\documentclass[12pt, draftclsnofoot, onecolumn]{IEEEtran}
\usepackage{amsfonts,amsmath,amssymb,algorithm,algorithmic,bm}
\usepackage{psfrag,graphicx,epsfig,cite}
\usepackage{color}
\usepackage{subfigure}
\usepackage[percent]{overpic}
\usepackage{multirow}
\usepackage{autobreak}
\usepackage{hyperref}
\usepackage{amsthm}
\usepackage{makecell}
\hypersetup{hidelinks}

\def \beqi{\begin{IEEEeqnarray}{rcl}\IEEEyesnumber}
\def \eeqi{\end{IEEEeqnarray}}

\def \bmat{\begin{bmatrix}}
\def \emat{\end{bmatrix}}

\newcommand{\blue}{\textcolor{black}}

\begin{document}

\title{Deep Reinforcement Learning for Distributed Dynamic Coordinated Beamforming in Massive MIMO Cellular Networks} 


\author{Jungang Ge, Ying-Chang Liang,~\IEEEmembership{Fellow,~IEEE}, Liao Zhang,\\ Ruizhe Long, and Sumei Sun,~\IEEEmembership{Fellow,~IEEE} 

\thanks{This work has been submitted to the IEEE for possible publication. Copyright may be transferred without notice, after which this version may no longer be accessible.}
\thanks{J. Ge, L. Zhang, and R. Long are with the National Key Laboratory of Science and Technology on Communications, and the Center for Intelligent Networking and Communications (CINC), University of Electronic Science and Technology of China (UESTC), Chengdu 611731, China (e-mail: {gejungang@std.uestc.edu.cn; zhangliao@std.uestc.edu.cn; ruizhelong@gmail.com}).
}
\thanks{Y.-C. Liang and S. Sun are with the Institute for Infocomm Research, Agency for Science, Technology and Research, Singapore 138632 (e-mail: {liangyc@ieee.org; sunsm@i2r.a-star.edu.sg}).}
}

\maketitle
\vspace{-1cm}
\begin{abstract}
To accommodate the explosive wireless traffics, massive multiple-input multiple-output (MIMO) is regarded as one of the key enabling technologies for next-generation communication systems. In massive MIMO cellular networks, coordinated beamforming (CBF), which jointly designs the beamformers of multiple base stations (BSs), is an efficient method to enhance the network performance. In this paper, we investigate the sum rate maximization problem in a massive MIMO mobile cellular network, where in each cell a multi-antenna BS serves multiple mobile users simultaneously via downlink beamforming. Although existing optimization-based CBF algorithms can provide near-optimal solutions, they require realtime and global channel state information (CSI), in addition to their high computation complexity. It is almost impossible to apply them in practical wireless networks, especially highly dynamic mobile cellular networks. Motivated by this, we propose a deep reinforcement learning based distributed dynamic coordinated beamforming (DDCBF) framework, which enables each BS to determine the beamformers with only local CSI and some historical information from other BSs.Besides, the beamformers can be calculated with a considerably lower computational complexity by exploiting neural networks and expert knowledge, i.e., a solution structure observed from the iterative procedure of the weighted minimum mean square error (WMMSE) algorithm. Moreover, we provide extensive numerical simulations to validate the effectiveness of the proposed DRL-based approach. With lower computational complexity and less required information, the results show that the proposed approach 
can achieve comparable performance to the centralized iterative optimization algorithms.

\end{abstract}

\section{Introduction}\label{sec:intro}

As wireless traffics grow dramatically in the modern digital society, multiple antennas have been widely deployed to improve the performance of communication systems since the third generation (3G) mobile networks \cite{adjoudani2003prototype}. Massive multiple-input multiple-output (MIMO), which exploits a large number of antennas to realize high-performance communication systems, is regarded as one of the most enthralling technologies for fifth-generation (5G) and future sixth-generation (6G) mobile networks \cite{chataut2020massive, marzetta2016fundamentals, liang2020dynamic, you2021towards, chen2020vision}.

With massive MIMO, higher array gains and more efficient interference suppression can be realized via various beamforming methods, e.g., the maximum ratio transmission (MRT) and zero-forcing (ZF) approaches. In massive MIMO cellular networks, coordinated beamforming (CBF) techniques are proposed to enhance the network performance by jointly designing the beamformers of all BSs, and several well-known optimization-based algorithms are developed to obtain near-optimal beamformers \cite{shi2011iteratively,shen2018fractional}. In \cite{shi2011iteratively}, a weighted sum-rate maximization problem is shown to be equivalent to a weighted sum mean square error minimization problem and then can be solved by the iterative weighted minimum mean square error (WMMSE) algorithm. Besides, as the weighted sum-rate maximization problem belongs to the family of optimization problems including ratio terms, it can also be solved by the fractional programming (FP) method \cite{shen2018fractional}. \blue{The equivalence between the WMMSE algorithm and the closed-form FP algorithm is also proved in \cite{shen2018fractional}, and it is shown that the direct FP algorithm can achieve a higher performance in comparison with the closed-form FP (WMMSE) algorithm.} Although these optimization-based algorithms can provide near-optimal solutions, they require real-time and global channel state information (CSI), and an iterative optimization procedure with high computational complexity. The execution of these algorithms may have to involve a central controller, which collects the real-time global CSI, executes the iterative optimization procedure, and sends back the obtained beamformers to all BSs. The communication overhead to share the CSI and the beamformer between the central controller and the BSs, therefore, will incur additional costs. In practical wireless networks, especially the highly dynamic mobile networks with fast-changing CSI, the BSs have to adapt their beamformers to the time-varying channels. In such cases, both the global CSI and the corresponding beamformers will have to be exchanged between the central controller and the BS more frequently, otherwise the performance will deteriorate. Besides, the large number of antennas lead to large dimensional optimization problems, which inevitably make the computational complexity of these algorithms very high. \blue{In \cite{choi2012distributed}, the authors decouple the weighted sum rate maximization problem into distributed subproblems of only local CSI and propose a zero-gradient based distributed CBF approach to eliminate the global CSI requirement. The local optimum with comparable performance to the centralized optimization algorithms can be obtained by letting each BS execute an iterative optimization procedure, which is still of high computational complexity. In \cite{zhao2023rethinking}, a reduced-WMMSE approach is proposed by leveraging a low-dimensional subspace property of the stationary point, and the computational complexity in each iteration is reduced as compared to the original WMMSE algorithm. However, the proposed approach is developed for single-cell scenarios and can not be directly extended to multi-cell scenarios. In addition, the global CSI requirement and the time-consuming iterative optimization process are also not alleviated.} As a result, despite the notable performance improvement, CBF remains practically challenging in a dynamic massive MIMO cellular network.

Recently, the exploitation of machine learning (ML) techniques, e.g., deep learning (DL) and deep reinforcement learning (DRL), provides us with alternative approaches to solve conventional communication problems \cite{sun2018learning,luong2019applications, he2020model}. These approaches have shown competitive advantages in computational complexity and amount of required information over the conventional optimization-based methods. Particularly, many machine learning based beamforming optimization approaches are investigated in conventional cellular communication systems. \blue{In \cite{xia2019deep, kim2020deep, kim2022bipartite}, the DL-based approaches exploiting expert knowledge, i.e., the known structure of optimal solutions, are studied for the beamforming optimization in a single-cell MU-MISO downlink system. Particularly, the exploitation of expert knowledge can improve the performance of the DL-based approaches \cite{kim2020deep}. In \cite{kim2022bipartite}, a bipartite graph neural network based approach is further developed to realize a scalable DL-based solution.} In \cite{schynol2022coordinated}, a deep unrolling approach is proposed to reduce the number of required iterations by unfolding the iterative optimization procedures as graph neural networks, however, it still requires the real-time global CSI and some iterations to obtain near-optimal beamformers. In \cite{mismar2020deep, ge2020deep}, the joint beamforming, power control and interference coordination problem in cellular networks is investigated. Specifically, the authors in \cite{mismar2020deep} consider a two-cell network and propose a centralized DRL-based approach, where only a single agent is employed to control the beamformers and transmit power of both the two cells. For the more general multi-cell networks, a distributed beamforming coordination approach based on multi-agent DRL is proposed in \cite{ge2020deep}. It is worth noting that the DRL-based schemes in \cite{mismar2020deep, ge2020deep} are designed for cellular networks with only a single user per cell, and thus can not be directly extended to more general multi-cell multi-user cellular networks. Moreover, the codebook-based method is adopted in \cite{mismar2020deep, ge2020deep}, where the optimal beamformers can only be selected from a predefined set of available beamformers. As such, when the channel characteristics are more complex, e.g., the channels in a lower frequency band instead of the millimeter wave band considered in \cite{mismar2020deep}, the optimal beamformers may not fall into the predefined set, and it is almost impossible to obtain the optimal beamformers with the codebook-based method due to the mismatch between the channel characteristics and the codebook.

Inspired by the huge potential of ML-based beamforming optimization approaches, we propose a DRL-based distributed dynamic coordinated beamforming (DDCBF) framework for a massive MIMO mobile cellular network. Specifically, a known solution structure from the WMMSE (closed-form FP) algorithm, which can be regarded as expert knowledge for CBF problems in multi-cell scenarios, is leveraged in the proposed approach. Besides, the proposed approach enables each BS to obtain the optimal beamformers with only local CSI and some historical information from other BSs, and the computational complexity is considerably reduced. Our main contributions and the advantages of the proposed DRL-based DDCBF framework are summarized as follows.
\blue{
\begin{itemize}
    \item As the iterative optimization-based algorithms suffer from strict CSI requirement and high computational complexity in massive MIMO mobile cellular networks, we propose a DRL-based DDCBF framework. Specifically, each BS can obtain the near-optimal beamformers with local CSI and some historical information from other BSs, hence alleviating the requirement for real-time global CSI. With the exploitation of neural networks, the computational complexity to calculate the beamformers is substantially reduced, and therefore each BS can adjust the beamformers quickly once the channel varies. 
    \item We utilize expert knowledge, i.e., a known solution structure that can be observed from the iterative procedures of the centralized optimization-based algorithms, to design the action space of the BSs. This structure enables each BS to determine the downlink beamformers with much fewer parameters in contrast to directly outputting the large dimensional beamformers for multiple users. Moreover, different from the centralized optimization-based algorithms, the key parameters for determining the beamformers at each BS are not necessarily the same, which makes it more possible for the proposed approach to learn an optimal policy. 
    \item The proposed distributed reward function design, which can be regarded as a decomposition of the original sum rate maximization problem, allows each BS to optimize the policy by maximizing a distributed reward. Incorporating with the local replay buffer, the training process can be realized in a decentralized manner, thus eliminating the large overhead of jointly training all the agents.
    \item The effectiveness of the proposed DRL-based approach is demonstrated via extensive simulations, where the quasi deterministic radio channel generator (QuaDRiGa) is employed to simulate a massive MIMO mobile cellular network under the three-dimensional urban macro cell (3D-UMa) scenario. With less required information and considerably lower computational complexity, the results show that the proposed DDCBF framework can outperform the iterative closed-form optimization algorithms and achieve a close performance to the state-of-the-art upper bound algorithm. 
\end{itemize}
}

The remainder of this paper is organized as follows. Section \ref{sec:sys_model} illustrates the system model, formulates a dynamic coordination beamforming problem, and elaborates on the known solution structure. In Section \ref{sec:DRL4DDIC}, the basics of DRL are introduced, a DRL-based DDCBF framework is presented, and the required information exchange procedure is discussed. In Section \ref{sec:sim}, we provide extensive numerical simulations to evaluate the proposed approach. Finally, Section \ref{sec:conclusion} concludes this paper.

\emph{Notations:} The notations in this paper are as follows. The scalars, column vectors and matrices are denoted by lowercase, bold lowercase and bold uppercase symbols (e.g., $x$, $\mathbf{x}$ and $\mathbf{X}$), respectively. $x^{\dag}$ denotes the conjugate of $x$. $\mathbf{x}^H$ and $\mathbf{X}^H$ respectively denote the Hermitian of $\mathbf{x}$ and $\mathbf{X}$. $|\cdot|$ and $\|\cdot\|$ are the absolute value and Euclidean norm operators. In particular, $|\mathcal{X}|$ is the cardinality of $\mathcal{X}$ with $\mathcal{X}$ denoting the set. $x\sim\mathcal{CN}(0, \sigma^2)$ means that $x$ is a complex Gaussian random variable with zero mean and variance $\sigma^2$. $x\sim\mathcal{N}(0, \sigma^2)$ means that $x$ is a real Gaussian random variable with zero mean and variance $\sigma^2$.


\section{Dynamic Coordinated Beamforming and Known Solution Structure}\label{sec:sys_model}

\subsection{System Model}\label{subsec:sig_model}

\begin{figure}[!htbp]
    \begin{center}
    \epsfxsize=0.5\linewidth
    \epsffile{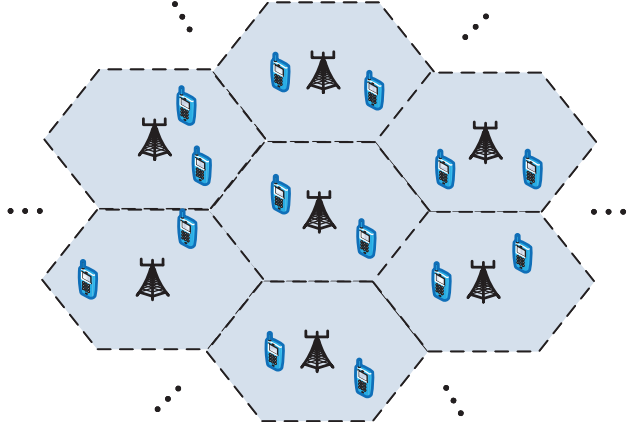}
    \caption{The considered massive MIMO cellular network.}\label{fig:sys_model}
    \end{center}
\end{figure}

\blue{As shown in Fig. \ref{fig:sys_model}, we consider a time division duplex (TDD) massive MIMO cellular network with $N$ cells, in each of which a centrally located base station (BS) serves $K$ single-antenna user equipments (UEs).}
In particular, we denote $\mathcal{B}=\{1, \cdots, N\}$ the set of the BSs, and each BS is equipped with a uniform rectangular array (URA) of $M=M_1\times M_2$ antennas. \blue{Besides, BS $m$ can obtain the local CSI \cite{park2012new}, i.e., $\mathbf{h}_{m,n,k}$($\forall n,k$), by utilizing the channel reciprocity, where $\mathbf{h}_{m,n,k}$ denotes the downlink channel between BS $m$ and UE $k$ in cell $n$.} With downlink beamforming, each BS serves its $K$ UEs simultaneously in the same frequency band. Let $\mathbf{w}_{n,k}$ denote the beamformer for UE $k$ in cell $n$, the received signal of UE $k$ in cell $n$ is
\begin{align}\label{eq:recesigynk}
    y_{n,k}=\underbrace{\mathbf{h}_{n,n,k}^H\mathbf{w}_{n,k}s_{n,k}}_{\text{desired signal}}+\underbrace{\sum_{j=1,j\neq k}^{K}\mathbf{h}_{n,n,k}^H\mathbf{w}_{n,j}s_{n,j}}_{\text{intra-cell interference}}
    +\underbrace{\sum_{l=1,l\neq n}^{N}\sum_{j=1}^{K}\mathbf{h}_{l,n,k}^H\mathbf{w}_{l,j}s_{l,j}}_{\text{inter-cell interference}} + u_{n,k},
\end{align}
where $s_{n,k}\sim\mathcal{CN}(0, 1)$ denotes the transmitted symbol for UE $k$ in cell $n$, $u_{n,k}\sim \mathcal{CN}(0, \sigma_u^2)$ denotes the additive white Gaussian noise (AWGN). Therefore, the signal-to-interference-plus-noise ratio (SINR) of UE $k$ in cell $n$ is
\begin{equation}\label{eq:SINRnk}
    \gamma_{n,k}=\frac{|\mathbf{h}_{n,n,k}^H\mathbf{w}_{n,k}|^2}{\sum_{j=1,j\neq k}^{K}|\mathbf{h}_{n,n,k}^H\mathbf{w}_{n,j}|^2+\sum_{l=1,l\neq n}^{N}\sum_{j=1}^{K}|\mathbf{h}_{l,n,k}^H\mathbf{w}_{l,j}|^2+\sigma_u^2}.
\end{equation}
Further, the achievable rate of UE $k$ in cell $n$ can be calculated as
\begin{equation}\label{eq:SEnk}
    R_{n,k}=\log_2\left(1+\gamma_{n,k}\right).
\end{equation}

\subsection{Dynamic Coordinated Beamforming}\label{subsec:ch_model}

\begin{figure}[!t]
    \begin{center}
        \psfrag{a}[cc][cc][1][0]{$\mathbf{h}(t-1)$}
        \psfrag{e}[cc][cc][1][0]{$\mathbf{h}(t)$}
        \psfrag{c}[cc][cc][1][0]{$\mathbf{h}(t+1)$}
        \epsfxsize=0.5\linewidth
        \epsffile{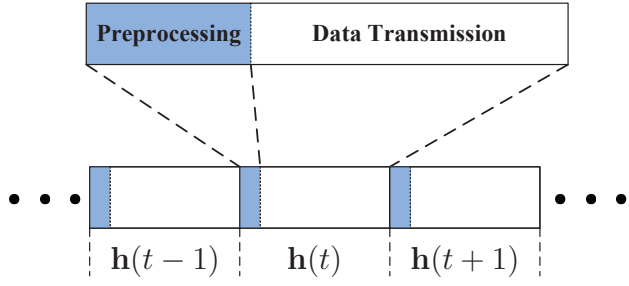}
        \caption{The proposed downlink transmission frame structure for the dynamic coordinated beamforming problem.}\label{fig:transmission_frame}
    \end{center}
\end{figure}

More practically, we consider a flat block-fading channel model under the three-dimensional urban macro cell (3D-UMa) scenario \cite{3gpp38901}, \blue{where the channel coefficients vary between different time slots and are temporally correlated because the movements of the mobile users are continuous}. As such, each BS has to accommodate the downlink beamformers to the time-variant channels. Considering that the synchronization is well realized, we propose a downlink transmission frame structure as depicted in Fig. \ref{fig:transmission_frame}. In particular, each frame is divided into two phases, namely, the preprocessing phase and the data transmission phase. In the preprocessing phase, the BS collects the necessary information and obtains the beamformers with a given beamforming optimization method. In the data transmission phase, the BS transmits data to its serving users with the determined beamformers.

Without loss of generality, we consider to enhance the sum rate performance of the considered mobile cellular network and formulate the following dynamic CBF problem as 
\begin{align}\label{eq:obj_func}
	\max \limits_{\mathbf{w}_{n,k}(t)} & \sum_{n=1}^{N}\sum_{k=1}^{K}R_{n,k}(t),\\
	{s.t.}\quad &\sum_{k=1}^{K}\|\mathbf{w}_{n,k}(t)\|^2\le{P_{\rm max}},\ \forall n, \nonumber
\end{align}
where $\mathbf{w}_{n,k}(t)$ and $R_{n,k}(t)$ denote the corresponding variables in time slot $t$, and $P_{\rm max}$ is the maximum transmit power budget of each BS. Further, the beamformer $\mathbf{w}_{n,k}(t)$ can be decomposed as
\begin{equation}\label{eq:bf_decom}
  \mathbf{w}_{n,k}(t)=\sqrt{p_{n,k}(t)}\bar{\mathbf{w}}_{n,k}(t),
\end{equation}
where $p_{n,k}(t)=\|\mathbf{w}_{n,k}(t)\|^2$ denotes the allocated power for UE $k$ in cell $n$ \cite{nguyen2014mmse, ge2020deep}, and $\bar{\mathbf{w}}_{n,k}(t)$ is the normalized beamformer. Note that arbitrary solutions to \eqref{eq:obj_func} can be decomposed as \eqref{eq:bf_decom}, by substituting \eqref{eq:bf_decom} to \eqref{eq:recesigynk}--\eqref{eq:SEnk}, \eqref{eq:obj_func} can be equivalently rewritten as
\begin{align}\label{eq:obj_func_PA}
	\max \limits_{\bar{\mathbf{w}}_{n,k}(t),\ p_{n,k}(t)} & \sum_{n=1}^{N}\sum_{k=1}^{K}R_{n,k}(t),\\
	{s.t.}\quad &\sum_{k=1}^{K}p_{n,k}(t)\le{P_{\rm max}},\ \forall n, \nonumber\\
    &\|\bar{\mathbf{w}}_{n,k}(t)\|^2=1,\  \forall n, k.\nonumber
\end{align}
Obviously, problem \eqref{eq:obj_func} is NP-hard and non-convex, and it is quite challenging to find the optimal solutions. Various well-known iterative optimization-based algorithms, e.g., the WMMSE algorithm \cite{shi2011iteratively} and the FP-based algorithm \cite{shen2018fractional} have been proposed to obtain the near-optimal solutions. However, these algorithms are of high computational complexity and require real-time global CSI. This means that every time the channel changes, we have to collect the required information and re-run these algorithms very quickly. In the considered dynamic massive MIMO cellular network, it is almost impossible to meet these requirements, and therefore these algorithms are actually not practically viable.

\subsection{Known Solution Structure}

Although the optimization-based algorithms above are not applicable in the practical dynamic wireless environment, their solutions, especially the iterative optimization procedures, can provide us with expert knowledge for the considered CBF problems. Specifically, the pseudo-code of the WMMSE algorithm for \eqref{eq:obj_func} is shown in Algorithm \ref{Alg:WMMSE}, where $u_{n,k}$'s and $v_{n,k}$'s can be regarded as the auxiliary variables introduced in the iterative process. Noting that $u_{n,k}$'s are complex and $v_{n,k}$'s are real, $u_{m,j}v_{m,j}u_{m,j}^{\dag}$ is actually a real variable. Let $\alpha_{m,j}\triangleq u_{m,j}v_{m,j}u_{m,j}^{\dag}$, the final step to obtain $\mathbf{w}_{n,k}$ in Algorithm \ref{Alg:WMMSE} can be rewritten as
\begin{equation*}
    \mathbf{w}_{n,k}^{\text{WMMSE}}\triangleq\left(\sum_{(m,j)}\alpha_{m,j}\mathbf{h}_{n,m,j}\mathbf{h}_{n,m,j}^H+\mu_{n}\mathbf{I}\right)^{-1}\mathbf{h}_{n,n,k}u_{n,k}v_{n,k}.
\end{equation*}
We then obtain the normalized beamformer as
\begin{equation}\label{eq:opt_beam}
    \bar{\mathbf{w}}_{n,k}=\frac{\left(\sum\limits_{(m,j)}\alpha_{m,j}\mathbf{h}_{n,m,j}\mathbf{h}_{n,m,j}^H+\mu_{n}\mathbf{I}\right)^{-1}\mathbf{h}_{n,n,k}}{\left\|\left(\sum\limits_{(m,j)}\alpha_{m,j}\mathbf{h}_{n,m,j}\mathbf{h}_{n,m,j}^H+\mu_{n}\mathbf{I}\right)^{-1}\mathbf{h}_{n,n,k}\right\|}.
\end{equation}
With \eqref{eq:bf_decom} and \eqref{eq:opt_beam}, the beamformer $\mathbf{w}_{n,k}=\sqrt{p_{n,k}}\bar{\mathbf{w}}_{n,k}$ can be determined by the local CSI, $\alpha_{m,j}$'s, $\mu_{n}$, and $p_{n,k}$. It is worth noting that this known solution structure can also be observed from the closed-form FP algorithm, since the equivalence between the closed-form FP algorithm and the WMMSE algorithm is illustrated in \cite{shen2018fractional}.

Furthermore, the known solution structure provides us with more insights into basic principles of the CBF approaches. Firstly, the normalized beamformer in \eqref{eq:opt_beam} actually maximizes a Rayleigh quotient as
\begin{align}\label{eq:rayleigh_quotient}
    \mathbf{w}_{n,k}^{\text{RQ}} & = \arg\max_{\mathbf{w}_{n,k}} \frac{\mathbf{w}_{n,k}^H\mathbf{h}_{n,n,k}\mathbf{h}_{n,n,k}^H\mathbf{w}_{n,k}}{\mathbf{w}_{n,k}^H\left(\sum\limits_{(m,j)}\alpha_{m,j}\mathbf{h}_{n,m,j}\mathbf{h}_{n,m,j}^H+\mu_{n}\mathbf{I}\right)\mathbf{w}_{n,k}}\\
    \label{eq:wslnr}
    &= \arg\max_{\mathbf{w}_{n,k}} \frac{|\mathbf{h}_{n,n,k}^H\mathbf{w}_{n,k}|^2}{\sum\limits_{(m,j)}\alpha_{m,j}|\mathbf{h}_{n,m,j}^H\mathbf{w}_{n,k}|^2+\mu_{n}\|\mathbf{w}_{n,k}\|^2}.
\end{align}
Letting $\|\mathbf{w}_{n,k}^{\text{RQ}}\|=1$, we can obtain the normalized beamformer in \eqref{eq:opt_beam} by solving \eqref{eq:wslnr}. The principle in \eqref{eq:wslnr} is similar to a conventional beamforming method which exploits local CSI to realize interference management, namely, maximum signal-to-leakage-and-noise ratio (SLNR) beamforming \cite{sadek2007leakage}. According to the maximum SLNR (Max-SLNR, also denoted by MSLNR in this paper) method, the beamformer is calculated by
\begin{equation}\label{eq:mslnr_beamformer}
    \mathbf{w}_{n,k}^{\rm MSLNR} = \arg\max_{\mathbf{w}_{n,k}} \frac{|\mathbf{h}_{n,n,k}^H\mathbf{w}_{n,k}|^2}{\sum\limits_{(m,j)\neq (n,k)}|\mathbf{h}_{n,m,j}^H\mathbf{w}_{n,k}|^2+\sigma_u^2}.
\end{equation}
Comparing \eqref{eq:wslnr} with \eqref{eq:mslnr_beamformer}, the normalized beamformer in \eqref{eq:opt_beam} can be regarded as a more general SLNR-based beamformer structure. Particularly, $\alpha_{m,j}$ can be seen as a weight for the interference leakage power to user $j$ in cell $m$ when calculating the beamformer of user $k$ in cell $n$, and $\mu_{n}$ can be regarded as a scaling factor for the AWGN. In the context, we refer to $\alpha_{m,j}$'s and $\mu_n$'s as the interference leakage control factor (ILCF) and background noise control factor (BNCF), respectively. Obviously, \eqref{eq:opt_beam} provides us with a more general beamformer structure to realize efficient interference management, which accounts for the impacts of both the AWGN and the interference leakage power to other users. Besides, \eqref{eq:opt_beam} can also represent most of the conventional beamforming methods. For example, the maximum ratio transmission (MRT) beamformer, i.e., $\bar{\mathbf{w}}_{n,k}^{\rm MRT} = {\mathbf{h}_{n,n,k}}/{\left\|\mathbf{h}_{n,n,k}\right\|}$, can be written as the beamformer structure with $\alpha_{m,j}=0$ and arbitrary $\mu_n$. The zero forcing (ZF) beamformer can also be interpreted as a beamformer that maximizes the signal-to-leakage ratio by letting the interference leakage be almost $0$. Through this beamformer structure, various beamformers can be calculated with different ILCFs and BNCFs to realize different interference management schemes. Hence, we can infer that there exist the optimal ILCFs and BNCFs providing optimal solutions to problem \eqref{eq:obj_func}. More importantly, the number of parameters to determine is mainly related to the number of users, which is much smaller than the dimension of the original beamforming matrix to optimize in massive MIMO cellular networks. Thus, the known solution structure allows us to obtain the optimal beamformers by determining much fewer parameters as compared to the schemes directly optimizing the large dimensional beamformers.

\begin{algorithm*}[hthp]
    \small{
    \caption{Pseudo-code of the WMMSE algorithm\label{Alg:WMMSE}}
    {
    \hspace*{\algorithmicindent} \textbf{Input:} Global CSI, i.e., $\mathbf{h}_{m,n,k}$, $\forall m,n,k$. \\
    \hspace*{\algorithmicindent} \textbf{Output:} Downlink beamformers, i.e., $\mathbf{w}_{n,k}$, $\forall n, k$.
    \begin{algorithmic}[1]
        \STATE Initialize $\mathbf{w}_{n,k}$ such that $\sum_{k=1}^{K}\|\mathbf{w}_{n,k}(t)\|^2={P_{\rm max}}$, $\forall n, k$.
        \REPEAT
        \STATE $v_{n,k}'\leftarrow v_{n,k}$, $\forall n,k$.
        \STATE $u_{n,k}\leftarrow(\sum_{(m,j)}\mathbf{h}_{n,m,j}^H\mathbf{w}_{m,j}\mathbf{w}_{m,j}^H\mathbf{h}_{n,m,j}+\sigma^2)^{-1}\mathbf{h}_{n,n,k}^H\mathbf{w}_{n,k}$, $\forall n, k$.
        \STATE $v_{n,k}\leftarrow(1-u_{n,k}^{\dag}\mathbf{h}_{n,n,k}^H\mathbf{w}_{n,k})^{-1}$, $\forall n, k$.
        \STATE $\mathbf{w}_{n,k}\leftarrow(\sum_{(m,j)}\mathbf{h}_{n,m,j}u_{m,j}v_{m,j}u_{m,j}^{\dag}\mathbf{h}_{n,m,j}^H+\mu_{n}\mathbf{I})^{-1}\mathbf{h}_{n,n,k}u_{n,k}v_{n,k}$, $\forall n, k$, where $\mu_{n}$ is an auxiliary variable for the maximum transmit power constraint and can be obtained via bisection search.
        \UNTIL $|\sum_{n,k}v_{n,k}-\sum_{n,k}v_{n,k}'|< \epsilon$, where $\epsilon$ represents the given stop condition.
        \STATE return $\mathbf{w}_{n,k}$, $\forall n, k$.
    \end{algorithmic}}
    }
\end{algorithm*}

\section{Deep Reinforcement Learning for Distributed Dynamic Coordinated Beamforming}\label{sec:DRL4DDIC}

\blue{
Firstly, we show that the dynamics of the considered system can be described with Markov decision process (MDP). As the delay for information exchange among BSs is non-negligible, it is fair to assume that the information from other cells is only available from the beginning of the next time slot. Thus, the downlink beamformers in the previous time slot will impact the collected information from other cells, such as the interference power information that we adopt as the elements of the agent's state space.} Then, we propose a DRL-based DDCBF framework, where the known solution structure is exploited to obtain the beamformers. The proposed approach enables each BS to determine the corresponding parameters for the optimal beamformers with only its local measurements and some historical information transferred from other BSs. As such, real-time global CSI is not required in the proposed framework. Besides, the computational complexity to calculate the optimal parameters is quite low since the parameters are directly obtained by neural networks, thus each BS can determine the beamformers very quickly once updated CSI is obtained.

\subsection{Preliminaries of DRL}\label{subsec:DRLbasics}

Deep reinforcement learning, which allows the intelligent agent to learn the policy through trial-and-error interactions with the environment, is an efficient machine learning method to tackle problems that can be formulated as \emph{Markov Decision Process} (MDP). An MDP problem can be described with a five-tuple, namely, $\langle\mathcal{S}, \mathcal{A}, \mathcal{T}, \mathcal{R}, \gamma\rangle$, which respectively represent the state space, the action space, the transition probabilities, the rewards, and the discount factor. Specifically, the interactions between the intelligent agent and the environment can be described as follows. At time step $t$, the agent in state $s(t)\in\mathcal{S}$ takes action $a(t)\in\mathcal{A}$ according to its policy $\pi$, then the agent's state changes from $s(t)$ to $s(t+1)\in\mathcal{S}$ with transition probability $P(s(t+1)|s(t), a(t))\in\mathcal{T}$, and finally the agent receives a reward $r(t)\in\mathcal{R}$ that evaluates the quality of the taken action. In general, the objective of the MDP problem is to obtain the optimal policy $\pi^*$ that maximizes a discounted cumulative reward, i.e., $G(t)=\sum_{\tau=0}^{\infty}\gamma^{\tau}r(t+\tau)$ with $\tau$ the step index relative to step $t$. With known environment dynamics, i.e., $\mathcal{T}$, the MDP problems can be solved by conventional dynamic programming methods. However, it is almost impossible to meet the requirements in the considered dynamic massive MIMO cellular network. The DRL-based approaches, which do not require accurate knowledge of the environment, provide us with promising solutions to the optimization problems in the complex wireless environment.

There are mainly two classes of DRL algorithms, namely, value-based and policy-based methods. The value-based methods, e.g., deep Q-learning \cite{mnih2015human}, learn the optimal policy by evaluating state-action values efficiently but can only deal with the MDP problems with discrete action space. Though the policy-based methods, e.g., policy gradient \cite{silver2014deterministic}, can tackle the MDP problems with continuous action space, however, they suffer from non-stationarity issues as the action can not be evaluated efficiently. Hence, the actor-critic framework, which exploits the advantages of both the value-based and policy-based methods, is proposed to realize continuous control with DRL. To be specific, the actor network provides a continuous mapping from state to action, and the critic network helps to train the actor network by evaluating the output action of the actor network. In our DDCBF problem, the beamforming parameters are obviously continuous values, and thus the action space is continuous. Therefore, we exploit one of the well-known actor-critic based DRL algorithms, namely, deep deterministic policy gradient (DDPG) \cite{lillicrap2016continuous}, to design the DDCBF framework.

The DDPG algorithm has two parts, i.e., actor and critic. Each part contains two neural networks, namely, the online network and the target network. The policy $\pi$ is represented by the (online) actor network, which can be denoted by a function as
\begin{equation}\label{eq:actor_function}
    a = \pi(s|\bm{\mu}),
\end{equation}
where $\bm{\mu}$ is the parameters of the actor network. On the other hand, the action is evaluated by the (online) critic network which can be represented by a Q value function $Q(s, a| \bm{\theta})$ with $\bm{\theta}$ the parameters of the critic network. The two target networks, i.e., the target actor network with parameters $\bm{\mu}'$ and the target critic network with parameters $\bm{\theta}'$, are exploited to realize a more stable offline training process. At each training step, $M_{b}$ experiences are sampled to form a mini-batch $\mathcal{M}$. Let $e=\langle s_e, a_e, r_e, s'_e\rangle\in\mathcal{M}$ denote an experience in the mini-batch, the loss function of the critic network is
\begin{equation}\label{eq:criticloss}
    L(\bm{\theta}) = \frac{1}{M_{b}}\sum_{e\in\mathcal{M}}\left(y_e^{\text{tar}}-Q(s_e, a_e|\bm{\theta})\right)^2,
\end{equation}
where $y_{e}^{\text{tar}}\triangleq r_{e}+\gamma Q'(s'_{e}, \pi(s'_{e}|\bm{\mu}')|\bm{\theta}')$, and the gradients to update the parameters of the actor network are
\begin{equation}\label{eq:actorgradient}
    \nabla_{\bm{\mu}} J(\bm{\mu}) =-\frac{1}{M_{b}}\sum_{e\in\mathcal{M}}\nabla_{a}Q(s, a|\bm{\theta})|_{s=s_{e}, a=a_{e}}
    \nabla_{\bm{\mu}}\pi(s|\bm{\mu})|_{s=s_{e}},
\end{equation}
with $J(\bm{\mu})$ the loss function of the actor network.

In addition, the DDPG algorithm can be executed by two parallel procedures: the online decision-making process and the offline training process. At time step $t$ of the online decision-making process, the agent obtains its state $s(t)$, determines an action by $a(t) = \pi(s(t)|\bm{\mu})+n_{a}(t)$ with $n_{a}(t)\sim\mathcal{N}(0, \sigma_a^2(t))$ the exploration noise, then receives a reward $r(t)$, and obtains its next state $s(t+1)$. Finally, the agent saves the experience, namely, $\langle s(t), a(t), r(t), s(t+1)\rangle$, into its experience replay memory $\mathcal{E}$, which is usually set with a first-input first-output (FIFO) queue. At each step of the offline training process, the agent samples a batch of $M_{b}$ experiences from its experience replay memory, and then trains the actor network and critic network respectively via minimizing the loss in \eqref{eq:criticloss} and the policy gradients in \eqref{eq:actorgradient}. Then, the parameters of the target actor network and the critic network are updated in a soft update manner, namely,
\begin{align}\label{eq:actorsoftupdate}
    \bm{\mu}'&\leftarrow \rho\bm{\mu} +(1-\rho)\bm{\mu}', \\\label{eq:criticsoftupdate}
    \bm{\theta}'&\leftarrow \rho\bm{\theta} +(1-\rho)\bm{\theta}',
\end{align}
where $\rho$ is a small constant.

\subsection{DRL-based DDCBF Framework}\label{subsec:DRL4DDCBF}


In a DDCBF framework, each BS should accommodate its beamformers to the time-variant wireless environment. Hence, each BS is a decision-maker, i.e., an intelligent agent, and the considered dynamic massive MIMO cellular network is a multi-agent system. However, the aforementioned DDPG algorithm is developed to solve the MDP problems in the single-agent system, and straightforwardly extending the algorithm into multi-agent systems may introduce several issues hindering the learning process, e.g., the non-stationarity and partial observability. These challenges can be solved by appropriately designing the state space and the reward function \cite{ge2020deep, nguyen2020deep}. In the following, we will elaborate on the proposed DRL-based DDCBF framework from the designs for the state space, the action space, and the reward function.

\noindent\textbf{State Space:}
For the intelligent agent, the state is its observation of the environment, which should include the information that helps the agent to identify the environment and make the right decision accordingly. In the multi-agent scenario, one agent should also acquire information from other agents to make the observed environment more stationary. Thus, we propose to design the agent's state with three types of information: the local BS's information, the interferer BSs' information, and the interfered UEs' information. \blue{It is worth noting that the latter two types of information are historical information acquired from other agents, and the historical information can also help the agent make decisions since the channel coefficients in adjacent time slots are temporally correlated.} Before describing the elements we designed for the agent's state, we give the following definitions for ease of notation in the context.

\blue{To reduce the number of state elements and the amount of information transferred from other BSs, we define $\mathbf{h}^{\rm c}$ as the compressed CSI of $\mathbf{h}\in\mathbb{C}^{M}$ on a uniform discrete Fourier transform (DFT) codebook $\mathbf{F}=[\mathbf{f}_1, \cdots, \mathbf{f}_C]\in \mathbb{C}^{M\times C}$, where $\mathbf{f}_c$ is given by
\begin{equation*}
\mathbf{f}_c = \frac{1}{\sqrt{M}}[1, e^{j\frac{2\pi}{C}c}, \cdots, e^{j\frac{2\pi}{C}(M-1)c}],
\end{equation*}
with $C$ the size of the codebook \cite{suh2017construction}.
Let $\mathbf{d}\triangleq\mathbf{F}^H\mathbf{h}=[d_1, d_2, \cdots, d_{C}]^T\in\mathbb{C}^{C}$, we sort the elements of $\mathbf{d}$ by their magnitudes, i.e., $|d_{c_{1}}|\geq|d_{c_{2}}|\geq\cdots\geq|d_{c_{C}}|$. Then, $\mathbf{h}^{\rm c}$ is given by $\mathbf{h}^{\rm c} = [c_{1}, d_{c_{1}}, c_{2}, d_{c_{2}}, \cdots, c_{N_{c}}, d_{c_{N_{c}}}]$, where $N_c$ denotes the compression factor.} Besides, we define the orthogonal measure matrix to indicate the spatial correlations of the multi-user channels $\mathbf{H}_{n}\triangleq[\mathbf{h}_{n,n,1}, \cdots, \mathbf{h}_{n,n,K}]$, and the $j$-th element of the $k$-th column of the orthogonal measure matrix $\mathbf{O}_{n}$ is
\begin{equation*}
    \mathbf{O}_{n}^{[j,k]}=\left[\frac{|\langle\mathbf{h}_{n,n,j}, \mathbf{h}_{n,n,k}\rangle|}{\|\mathbf{h}_{n,n,j}\|\|\mathbf{h}_{n,n,k}\|}\right]^2,
\end{equation*}
where $\langle\cdot, \cdot\rangle$ is the inner product operator. In addition, for UE $k$ in cell $n$, we denote the received signal power by $p^{\rm r}_{n,k}\triangleq p_{n,k}|\mathbf{h}_{n,n,k}^H\bar{\mathbf{w}}_{n,k}|^2$, the interference from BS $m$ by
\begin{equation*}
    \beta_{m,n,k}=\sum_{j=1\atop(m,j)\neq(n,k)}^{K}p_{m,j}|\mathbf{h}_{m,n,k}\bar{\mathbf{w}}_{m,j}|^2,
\end{equation*}
and the total interference plus noise by $\beta_{n,k} =\sum_{m=1}^{N}\beta_{m,n,k}+\sigma_u^2$.

With the notations above, we define the local information of BS $n$ in time slot $t$ as
\begin{align*}
    s^{\rm loc}_n(t)=&\left[\mathbf{O}_{n}(t),\mathbf{H}^{\rm c}_{n}(t), \mathbf{p}_{n}(t-1),R_{n,1}(t-1),\cdots, R_{n,K}(t-1), \mathbf{p}_{n}^{\rm r}(t-1), \bm{\beta}_{n}(t-1)\right],
\end{align*}
where $\mathbf{H}^{\rm c}_{n}(t)\triangleq[\mathbf{h}^{\rm c}_{n,n,1}(t), \cdots, \mathbf{h}^{\rm c}_{n,n,K}(t)]$ denote the compressed CSI of the UEs in cell $n$, $R_{n,1}(t-1),\cdots, R_{n,K}(t-1)$ are the achievable rates of the UEs in the previous slot,
$\mathbf{p}_{n}(t-1)\triangleq [p_{n,1}(t-1), \cdots, p_{n,K}(t-1)]$ denotes the power allocation in the previous slot, $\mathbf{p}_{n}^{\rm r}(t-1)\triangleq [p_{n,1}^{\rm r}(t-1), \cdots, p_{n,K}^{\rm r}(t-1)]$ is the received signal power of the UEs in the previous slot, and $\bm{\beta}_{n}(t-1)=[\beta_{n,1}(t-1), \cdots, \beta_{n,K}(t-1)]$ denotes the interference-plus-noise power of the UEs in the previous slot.

Moreover, to describe the information from other agents, we define the set of the interferer BSs of UE $k$ in cell $n$ as
\begin{equation}\label{eq:interfererBS}
\mathcal{I}^{\rm in}_{n,k}(t) = \left\{m\in\mathcal{B}| \beta_{m,n,k}(t)> \xi_{n,k}(t) \right\},
\end{equation}
and the set of the interfered UEs of cell $n$ as
\begin{equation}\label{eq:interferedUE}
\mathcal{I}^{\rm out}_{n}(t) = \left\{(m,j)\in\mathcal{U}| \beta_{n,m,j}(t)> \xi_{n}(t) \right\},
\end{equation}
where $\mathcal{U}\triangleq\{1,\cdots, N\}\times \{1,\cdots, K\}$ denotes the set of the indices of all the UEs, $\xi_{n,k}(t)$ and $\xi_{n}(t)$ are the thresholds that guarantee $|\mathcal{I}^{\rm in}_{n,k}(t)|=U$ and $|\mathcal{I}^{\rm out}_{n}(t)|=KU$. Further, we define the information from the interferer BSs of UE $k$ in cell $n$ as
\begin{align*}
    s^{\rm in}_{n,k}(t)=&\Big[\left[i,\mathbf{H}^{\rm c}_{i}(t-1), \mathbf{p}_{i}(t-1), \beta_{i,n,k}(t-1)\right]_{i\in\mathcal{I}^{\rm in}_{n,k}(t-1)}\Big],
\end{align*}
and the interferer BSs' information for agent $n$ as
\begin{equation*}
    s^{\rm in}_{n}(t)=[s^{\rm in}_{n,1}(t), \cdots, s^{\rm in}_{n,K}(t)].
\end{equation*}
Then, the interfered UEs' information is defined as
\begin{align*}
    s^{\rm out}_n(t)=\Big[\left[(m-1)K+j,R_{m,j}(t-1),\beta_{n,m,j}(t-1),\beta_{n,m,j}(t-1)/\beta_{m,j}(t-1)\right]_{(m,j)\in\mathcal{I}^{\rm out}_{n}(t-1)}\Big].
\end{align*}
Finally, the state of agent $n$ in time slot $t$ is
\begin{align*}
    s_n(t)=\left[s^{\rm loc}_n(t), s^{\rm in}_n(t), s^{\rm out}_n(t)\right].
\end{align*}

\noindent\textbf{Action Space:}
In regard to applying machine learning techniques to the beamformer optimization problems, the most intuitive idea is to directly output the beamformers by the deep neural networks. However, the beamformers are usually multi-dimensional complex vectors with given power constraints, thus it is quite difficult for the DRL agent to find the optimal beamformers via limited trial-and-error interactions with the environment, especially in the considered massive MIMO cellular network where the BSs are equipped with massive antennas. Besides, the codebook-based methods \cite{ge2020deep, mismar2020deep}, in which the beamformers are selected from a predefined codebook, only work well in some special scenarios. For instance, the DFT-based codebook can be utilized to design the beamforming approaches in millimeter wave communication systems, where the channels are usually assumed with a dominant line-of-sight component. When the channel characteristics are more complex, e.g., the channels in a lower frequency band under urban scenarios, it is almost impossible to obtain the optimal beamformers due to the mismatch between the channel characteristics and the beamforming codebook. Hence, a more general method to obtain the beamformers is required for designing the DRL-based beamforming optimization approaches.

With the exploitation of expert knowledge, the action space of BS $n$ ($n\in\mathcal{B}$) is given by
\begin{equation}\label{eq:ddpgaction}
    a_{n} = [q_{n,1}, \cdots, q_{n,K}, q_n^{\rm total}, \alpha_{n, 1,1}, \cdots, \alpha_{n,N,K}, \mu_{n}],
\end{equation}
where $q_{n,k}\in (0,1]$ with $\sum_{k=1}^{K}q_{n,k}=1$ denotes the ratio of the allocated transmit power for UE $k$ to the total transmit power of BS $n$, and $q^{\rm total}_{n}\in(0,1]$ is the ratio of the total transmit power of BS $n$ to the maximum transmit power. Besides, $\alpha_{n, 1,1}, \cdots, \alpha_{n,N,K}$ and $\mu_{n}$ are the ILCFs and BNCF for calculating the normalized beamformers. With the output action of BS $n$, we can finally obtain the allocated transmit power and the normalized beamformer of UE $k$ in cell $n$ by $p_{n,k}=P_{\rm max}q_n^{\rm total}q_{n,k}$ and \eqref{eq:opt_beam}, respectively. The corresponding beamformer for UE $k$ in cell $n$ can be recovered as $\mathbf{w}_{n,k}=\sqrt{p_{n,k}}\bar{\mathbf{w}}_{n,k}$. It is worth noting that the proposed design can significantly reduce the dimension of the action space, i.e., the number of required output ports of the actor network, as compared to intuitive designs outputting the beamformers directly. The lower dimensional action space also makes it possible for the BS to learn the optimal beamforming policy via limited trial-and-error interactions with the wireless environment.

\noindent\textbf{Reward Function:}
In this paper, the objective of each BS is to maximize the global sum rate via coordinated beamforming. From the viewpoint of game theory, the global sum rate usually cannot be maximized if each BS only tries to maximize the sum rate of its own users without considering the users in other cells. Therefore, we adopt a distributed reward function design, in which each agent's action is assessed by evaluating the impact of the taken action on both its serving UEs and the UEs in other cells. Motivated by the reward function designs adopted in \cite{nasir2019multi, ge2020deep,tan2020deep, zhao2019deep}, the reward function for agent $n$ is designed as
\begin{equation}\label{eq:reward_func}
    r_n(t)=\sum_{k=1}^{K}R_{n,k}(t) - \sum_{(m,j)\in\mathcal{I}^{\rm out}_{n}(t)}P_{m,j}(t),
\end{equation}
where
\begin{equation}\label{eq:penalty}
    P_{m,j}(t)=\log_{2}\left(1+\frac{p_{m,j}^{\rm r}(t)}{\beta_{m,j}(t)-\beta_{n,m,j}(t)}\right)-R_{m,j}(t).
\end{equation}
In \eqref{eq:reward_func}, the first term is the sum rate of the UEs in cell $n$, and the second term is a penalty for the caused interference to its interfered UEs. To be specific, the penalty actually quantifies the rate loss of the interfered UEs due to the interference caused by BS $n$, and the first term in \eqref{eq:penalty} is the achievable rate of user $j$ in cell $m$ if BS $n$ did not exist. Through this reward function, the BS can learn a policy that maximizes the sum rate of its serving UEs while trying to reduce the caused interference to the interfered UEs as well. Thus, this reward function can help each BS to realize efficient interference management and improve the network performance.

\begin{figure}[!t]
    \begin{center}
        \psfrag{0}[cr][cc][.8][0]{\rm soft update}
        \psfrag{1}[cl][cc][.8][0]{$s'_e$}
        \psfrag{2}[cr][cc][.8][0]{\rm $(s_e,a_e)$}
        \psfrag{3}[cr][cc][.8][0]{\rm soft update}
        \psfrag{4}[cr][cc][.8][0]{\rm $s'_e$}
        \psfrag{5}[lc][cc][.8][0]{$Q(s'_e,a'_e|\bm{\theta}'_n)$}
        \psfrag{6}[lc][cc][.8][0]{$Q(s_e,a_e|\bm{\theta}_n)$}
        \psfrag{7}[lr][cc][.8][0]{\rm update}
        \psfrag{8}[cc][cc][.8][0]{\rm $a'_e$}
        \psfrag{9}[cc][cc][.8][0]{\rm $r_e$}
        \psfrag{A}[cc][cc][.9][0]{\rm $s_n(t)$}
        \psfrag{B}[cc][cc][.8][0]{\rm Experience}
        \psfrag{C}[cc][cc][.9][0]{\rm $a_n(t)$}
        \psfrag{E}[cc][cc][.52][0]{\rm Experience}
        \psfrag{F}[cc][cc][.52][0]{\rm Replay Memory}
        \psfrag{G}[cc][cc][.9][0]{\rm Wireless Network Environment}
        \psfrag{M}[cc][cc][.9][0]{\rm BS $n$}
        \psfrag{j}[cc][cc][.8][0]{\rm BS $n$}
        \psfrag{H}[lc][cc][.8][0]{\rm Online Decision-making}
        \psfrag{K}[lc][cc][.8][0]{\rm Offline Training}
        \psfrag{S}[cc][cc][.7][0]{\rm Minibatch}
        \psfrag{R}[cc][cc][.45][0]{\rm Target Actor Network}
        \psfrag{P}[cc][cc][.6][0]{\rm Actor Network}
        \psfrag{L}[cc][cc][.8][0]{\rm Loss}
        \psfrag{N}[cc][cc][.8][0]{\rm Function}
        \psfrag{T}[cc][cc][.45][0]{\rm Target Critic Network}
        \psfrag{W}[cc][cc][.6][0]{\rm Critic Network}
        \psfrag{V}[cc][cc][.8][0]{\rm Optimizer}
        \psfrag{Y}[cc][cc][.6][0]{\rm Policy Gradient}
        \psfrag{Z}[cc][cc][.75][0]{\rm Add Noise}
        \epsfxsize=0.9\linewidth
        \epsffile{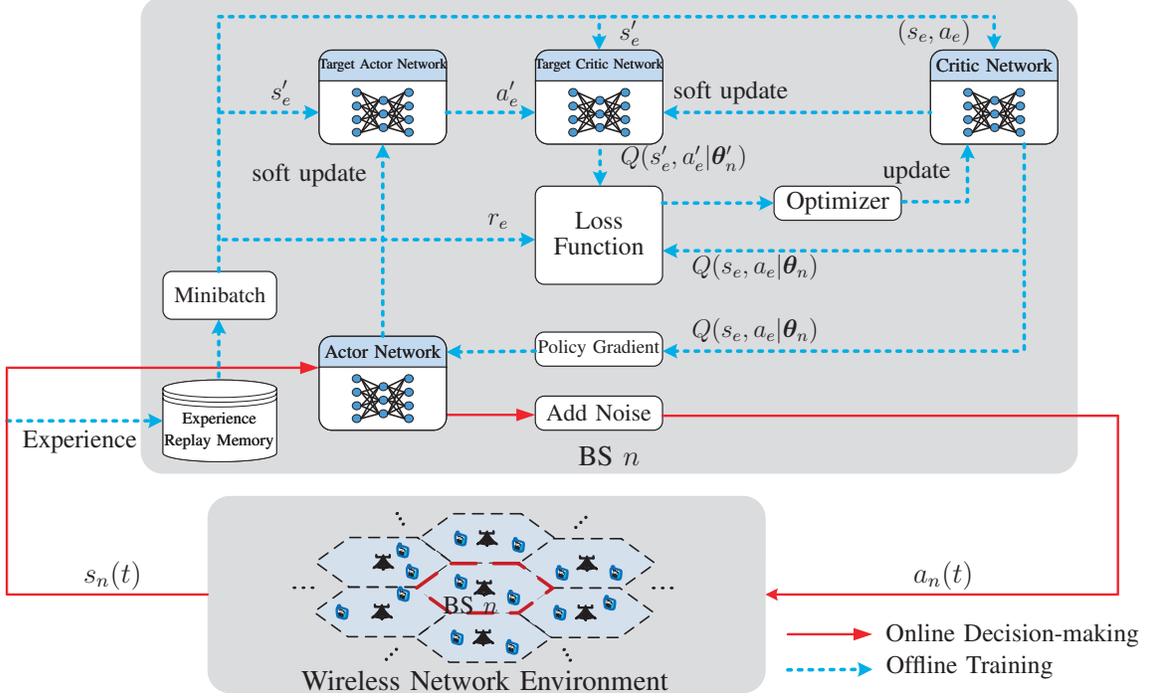}
        \caption{Illustration for the workflow of the proposed DRL-based DDCBF framework.}\label{fig:algorithm_illustration}
    \end{center}
\end{figure}

Finally, the workflow of the proposed DRL-based DDCBF framework is illustrated in Fig. \ref{fig:algorithm_illustration}. Specifically, the red solid lines represent the online decision-making process, the blue dashed lines show the offline training process, and the two procedures can be executed in a parallel manner. At time step $t$ of the online decision-making process, agent $n$ (BS $n$) obtains its state $s_n(t)$ by observing the wireless environment, then calculates the action with the actor network, and gets the final action $a_n(t)$ by adding a Gaussian exploration noise $n_a(t)\sim\mathcal{N}(0, \sigma_a^2(t))$. Finally, BS $n$ calculates the beamformers with the final action and the known solution structure. In the offline training procedure, the agent collects the trial-and-error experience, e.g., $\langle s_n(t), a_n(t), r_n(t), s_n(t+1)\rangle$, and saves the experience into its experience replay memory $\mathcal{E}_n$. At each training step, a mini-batch of $M_b$ experiences is sampled from the experience replay memory, letting $\mathcal{M}$ and $e=\langle s_e, a_e, r_e, s'_e\rangle\in\mathcal{M}$ denote the mini-batch and an experience of $\mathcal{M}$, the loss function for the critic network and the Q values for evaluating the actions, namely, \eqref{eq:criticloss} and $Q(s_e,a_e|\bm{\theta}_n)$, can be calculated. Then, the critic network and actor network are updated with the optimizer and the policy gradient in \eqref{eq:actorgradient}, respectively. Finally, the target actor network and the target critic network are updated in a soft manner as shown in \eqref{eq:actorsoftupdate} and \eqref{eq:criticsoftupdate}. In addition, the pseudo-code of the proposed DRL-based DDCBF framework is shown in \textbf{Algorithm \ref{alg:DRL4DDCBF}}.
\begin{algorithm}[!htbp]
    \small{
    \caption{Pseudo-code of the proposed DRL-based DDCBF framework}
    {
    \begin{algorithmic}[1]\label{alg:DRL4DDCBF}
    \STATE Agent $n$ randomly initializes the online actor network and critic network with weights $\bm{\mu}_n$ and $\bm{\theta}_n$, $\forall n \in \mathcal{B}$.
    
    \STATE Agent $n$ initializes the target actor network and the target critic network by $\bm{\mu}'_n\leftarrow \bm{\mu}_n$ and $\bm{\theta}'_n\leftarrow\bm{\theta}_n$, $\forall n \in \mathcal{B}$.
    
    \STATE Agent $n$ initializes the experience replay memory $\mathcal{E}_n$ with an FIFO queue of size $M_m$, $\forall n \in \mathcal{B}$.
     
    \STATE In time slot $t$ ($t\leq M_b$), agent $n$ takes action randomly and stores the experience $\langle s_n(t), a_n(t), s_n(t+1), r_n(t)\rangle$ into $\mathcal{E}_n$, $\forall n \in \mathcal{B}$.

    \REPEAT
    \STATE In time slot $t$ ($t>M_b$), agent $n$ obtains its state $s_n(t)$ by observing the wireless environment, $\forall n \in \mathcal{B}$.
    
    \STATE Agent $n$ gets an noised action by adding a Gaussian exploration noise $n_{a}(t)\sim\mathcal{N}(0, \sigma_a^2(t))$ on the output of the actor network, then clips the noised action to valid values to obtain the final action $a_n(t)$,  and finally update $\sigma_a(t+1)=\max(\sigma_a(t)/(1+\epsilon_{a}), \sigma_{a, \text{min}})$, $\forall n \in \mathcal{B}$.
    
    \STATE Agent $n$ executes the action $a_{n}(t)$, then receives an immediate reward $r_n(t)$ via \eqref{eq:reward_func}, $\forall n \in \mathcal{B}$.
    
    \STATE Agent $n$ obtains the next state $s_n(t+1)$, and stores the experience $\langle s_n(t), a_n(t), s_n(t+1), r_n(t)\rangle$ into $\mathcal{E}_n$, $\forall n \in \mathcal{B}$.
    
    \STATE Agent $n$ samples a mini-batch of $M_b$ experiences from $\mathcal{E}_n$, and respectively updates the actor network and the critic network via the policy gradients in \eqref{eq:actorgradient} and the loss function in \eqref{eq:criticloss}, with learning rates $\alpha_{\rm actor}$ and $\alpha_{\rm critic}$, $\forall n \in \mathcal{B}$.
    \STATE Agent $n$ softly updates the target actor network and the target critic network by \eqref{eq:actorsoftupdate} and \eqref{eq:criticsoftupdate}, $\forall n \in \mathcal{B}$.
    \UNTIL{convergence.}
    \end{algorithmic}}
    }
    \end{algorithm}

\subsection{Required Information Acquisition Procedure}

\begin{figure}[!tbp]
    \begin{center}
        \epsfxsize=0.5\linewidth
        \epsffile{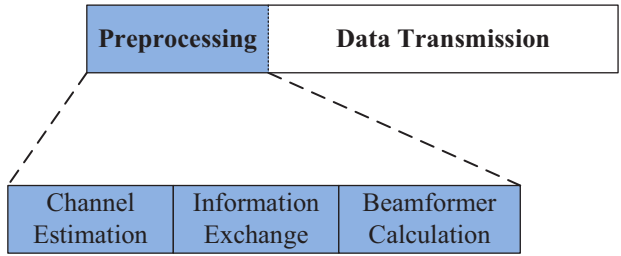}
        \caption{Three sub-phases of the preprocessing phase in the proposed DRL-based DDCBF framework.}\label{fig:pre_phase}
    \end{center}
\end{figure}

To realize the proposed DRL-based DDCBF framework, as the agent's state contains the local information and some historical information from other cells, an information acquisition procedure among the cells is required in the preprocessing phase of each time slot. Moreover, the local CSI, i.e., $\mathbf{h}_{n,m,j}$'s, which is necessary for recovering the beamformers from the agent's action, should also be estimated in the preprocessing phase. Therefore, the preprocessing phase can be divided into three sub-phases, which are shown in Fig. \ref{fig:pre_phase}. \blue{In the preprocessing phase, each BS first estimates local CSI by utilizing the channel reciprocity, then executes the information acquisition procedure, and finally determines the beamformers for its serving users.}

\blue{
In particular, the information acquisition procedure is as follows. With the obtained local CSI, the orthogonal measure matrix $\mathbf{O}_n$ and the compressed CSI $\mathbf{H}^{\rm c}_{n}$ can be calculated by BS $n$. Besides, noting that $\beta_{m,n,k}$ denotes the interference power from BS $m$ to UE $k$ in cell $n$, BS $m$ can calculate $\beta_{m,n,k}$ with $\mathbf{w}_{m,j}$ ($j=1, \cdots, K$) and $h_{m,n,k}$, which are all available at BS $m$. Then, each BS shares the interference-related information via the predefined interfaces for information exchange among the BSs, e.g., the X2 interface in LTE networks. Moreover, the information exchange procedure among BSs can be executed as follows. Firstly, each BS collects the interference-related information, namely, $\beta_{m,n,k}$'s. Then, each BS determines its interferer BSs and the interfered UEs, which are defined in \eqref{eq:interfererBS} and \eqref{eq:interferedUE}. Finally, each BS can obtain the required information, i.e., the information defined in $s^{\rm in}$ and $s^{\rm out}$.} Take BS $n$ as an example, the interferer BSs' information can be obtained by requesting the compressed CSI and the power allocation from BS $i\in\mathcal{I}^{\rm in}_{n,k}$ ($k=1,\cdots, K$). The interfered UEs' information can be also acquired by requesting the achievable rate, the interference power, and the total-interference-plus-noise power of the interfered UEs from the BS $m$ with $(m,j)\in\mathcal{I}^{\rm out}_{n}$. \blue{Besides, as the information transferred from other BSs is historical information, the information exchange process can be started in the previous time slot such that the inter-cell information becomes available at the beginning of the current time slot. As such, the delay of the inter-cell information exchange process in practical networks can be tackled.} Based on this information acquisition procedure, the proposed DRL-based DDCBF framework can be deployed in the considered dynamic massive MIMO cellular network.

\blue{\subsection{Discussions on the Handcraft Designs of the Proposed DRL-based Approach}}
\blue{
In the most of existing literature, including the work in this paper, the DRL-based approaches for wireless communications usually contain some handcraft designs providing promising performance. Beyond the common designs such as the historical action related information and achievable rates, there are actually some reasonable considerations in the proposed designs. With the fact that the performance gain of the CBF methods mainly comes from efficient interference management, the optimal solutions for ILCFs and BNCFs should depend on the interference-related information in the wireless environment, especially the considered system with multiple interfering cells. Hence, we involve the interference-related information from both the interferer BSs and interfered UEs of a given cell, to describe the interference environment of the cell. Moreover, the downlink beamformers of a BS also depend on the intra-cell channel orthogonality and the main directions of users' channels, hence adding the information into the state may also help the BS to learn a better policy. Besides, there are several similar designs in various wireless interference scenarios, e.g., mobile ad hoc networks \cite{nasir2019multi}, overlay D2D networks \cite{tan2020deep}, and symbiotic radio networks \cite{zhang2020intelligent}, where the effectiveness has been demonstrated. Therefore, these considerations and the core idea for these handcraft designs can be extended to other similar communication scenarios.}

\blue{
Despite the excellent performance of the heuristic designs, it may require time-consuming and trial-and-error-based numerical simulations to optimize the designs. Another promising approach, namely, ``learning to cooperate'' or ``learning to communicate'', allows the agents to optimize the inter-agent interaction rules autonomously by utilizing the feature extraction capability of the neural network \cite{lee2022artificial, kim2018learning, foerster2016learning, tung2021effective}. Specifically, each agent employs an additional neural network, which is referred to as message generating neural network (MGNN), to produce the messages for inter-agent information exchange based on its local information. Then, each BS can make decisions based on the local information the acquired messages from other agents, which is the same as illustrated in the proposed approach. However, the MGNNs should be trained together with the decision-making neural networks, and the neural networks of all BSs have to be trained jointly in a centralized manner. As a consequence, end-to-end training is required, and this will lead to a prohibitively large training overhead in the considered massive MIMO multicell system. Moreover, as the BSs optimize their policies during the interactions with the wireless environment in the DRL-based approach, the gradients have to be passed between agents in the back-propagation process \cite{foerster2016learning}, which incurs additional communication overhead for the back-haul networks. Obviously, the autonomous interaction rule optimization approach is of higher overhead in terms of both computation and communication, as compared to the handcraft designs proposed in our work where the neural networks of each BS can also be trained in a decentralized manner. Therefore, how to realize a tradeoff between time-consuming trial-and-error-based handcraft designs and a lower-overhead autonomous interaction rule optimization method may deserve more research in the future.}

\section{Numerical Results}\label{sec:sim}

In this section, we provide extensive numerical simulations to demonstrate the effectiveness of the proposed DRL-based DDCBF framework. Specifically, the proposed approach is evaluated by the sum achievable rate of all UEs, and we consider the following methods as benchmarks.
\begin{itemize}
    \item \textbf{MSLNR-EP:} The normalized beamformers are determined by max-SLNR method, and the equal power allocation is adopted, i.e., $p_{n,k}=P_{\text{max}}/K$.
    \item \textbf{MSLNR-DDPG:} The normalized beamformers are determined by max-SLNR method, and the power allocation is obtained by the proposed approach with the action space redesigned with only the power allocation related actions, i.e., $a_n=[q_{n,1}, \cdots, q_{n,K}, q_n^{\rm total}]$.
    \item \textbf{WMMSE:} The genie-aided WMMSE algorithm where we assume that the real-time global CSI is available and the computational time is negligible. The sum rate in each time slot is obtained with a single random initialization.
    \blue{
    \item \textbf{WMMSE-$N$RI:} The sum rate in each time slot is obtained with the largest sum rate achieved by the genie-aided WMMSE algorithm over $N$ random initializations.
    \item \textbf{Close-form FP:} The genie-aided closed-form FP algorithm \cite{shen2018fractional}, which is equivalent to the WMMSE algorithm. The sum rate in each time slot is obtained with a single random initialization.
    \item \textbf{Direct FP:} The genie-aided direct FP algorithm \cite{shen2018fractional}, where in each iteration the beamformers are updated by solving a convex optimization problem.
    \item \textbf{ZG:} An iterative distributed approach based on zero gradient condition, which operates with only local CSI \cite{choi2012distributed}.}
\end{itemize}

\subsection{Simulation Setup}

\begin{figure}[tbp]
    \begin{center}
    \epsfxsize=0.7\linewidth
    \epsffile{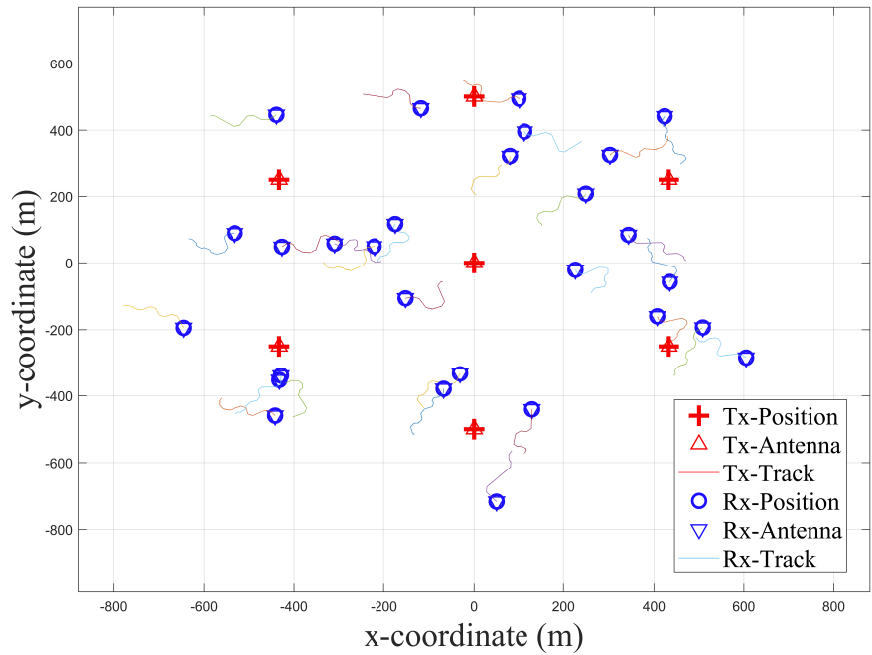}
    \caption{Visualization for the simulated scenario with $K=4$, generated by QuaDRiGa.}\label{fig:sim_topo}
    \end{center}
\end{figure}

In the simulations, we utilize the quasi deterministic radio channel generator (QuaDRiGa) \cite{jaeckel2014quadriga} to generate the channels in the considered massive MIMO mobile cellular network. With QuaDRiGa, the channel coefficients can be calculated by the ray-tracing technique, while the blocks and scatters in the wireless environment are randomly generated. Besides, the time variance of the channels, caused by the dynamics of the wireless environment, e.g., the movements of users, can also be tracked. The simulated scenario is shown in Fig. \ref{fig:sim_topo}, which is generated by QuaDRiGa. The red crosses denote the BSs' fixed positions, the blue circles denote the initial positions of the UEs, and the colored curves denote the tracks of the UEs. Particularly, we consider a  cellular network with $N=7$ cells, in each of which $K=4$ UEs are uniformly distributed. The cell radius, namely, half of the distance between two adjacent BSs, is set to $250$ m. The size of the URA at each BS is set as $M_1=M_2=8$, i.e., $M=64$. The time slot interval is set as $20$ ms. Furthermore, we set the system carrier frequency $f_c=2.6$ GHz, the speed of each UE $v=3$ km/h, the maximum transmit power budget $P_{\rm max}=38$ dBm, and the power of the AWGN $\sigma_u^2=-101$ dBm. Then, the channels are generated under the three-dimensional urban macro cell (3D-UMa) scenario according to 3GPP TR 38.901 \cite{3gpp38901}.

As for the proposed DRL-based DDCBF framework, the hyper-parameters are set as follows. \blue{To generate the DFT codebook for reducing the number of state elements and the amount of information transferred from other BSs, we set $C=128$ and $N_c=3$}. For the required inter-cell information exchange procedure, we set $U=4$, then we have $|\mathcal{I}^{\rm in}_{n,k}(t)|=U=4$ and $|\mathcal{I}^{\rm out}_{n}(t)|=KU=16$. For the adopted DDPG algorithm, the hidden neural networks of the actor and critic networks are all three-layer with $256$, $128$, and $64$ neurons, respectively. The experience replay memory at each BS is set as a FIFO queue with size $M_{m}=2000$ and the size of the mini-batch is set as $M_{b}=256$. The Adam optimizer is adopted for updating the critic network. Moreover, the learning rates for updating the actor network and the critic network are respectively set as $\alpha_{actor}=5\times 10^{-6}$ and $\alpha_{critic}=5\times 10^{-5}$. The small constant for softly updating the target networks is set as $\rho=0.01$. The discount factor for calculating the cumulative rewards is set as $\gamma=0.5$. The parameters of the action noise for exploration are set as $\sigma_a(0)=0.6$, $\epsilon_a=0.001$, and $\sigma_{a, \text{min}}=0.01$. We adopt ReLU as the activation function of the critic network and the hidden layers of the actor network. In particular, the sigmoid activation function is adopted in the output layer of the actor network, as we normalize all the elements of the action space, i.e., \eqref{eq:ddpgaction}, to the range $[0,1]$.

\subsection{Performance Comparison and Analysis}

\begin{figure}[tbp]
    \begin{center}
    \epsfxsize=0.5\linewidth
    \epsffile{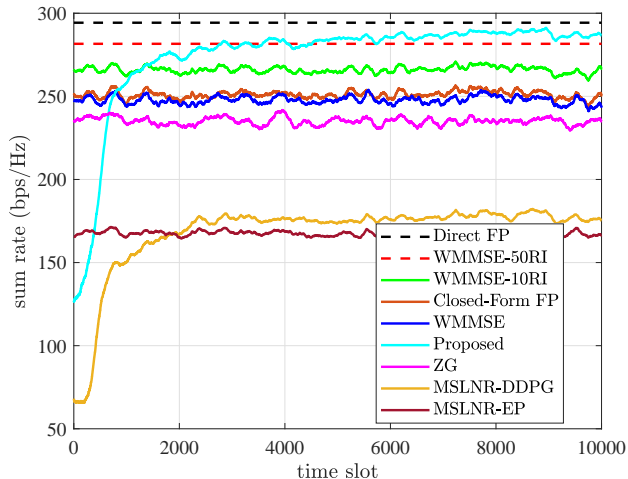}
    \caption{The convergence of the proposed DRL-based DDCBF framework in the simulated scenario. Each value is a moving average of the previous $200$ time slots.}\label{fig:perf_cmp}
    \end{center}
\end{figure}

Firstly, we validate the effectiveness of the proposed DRL-based DDCBF framework in comparison with the benchmarks. The convergence of the proposed approach is shown in Fig. \ref{fig:perf_cmp}. In particular, as the computational complexity to obtain the sum rates of $10000$ time slots is prohibitively large for the WMMSE-$50$RI and direct FP algorithms, we evaluate them by an average sum rate of $100$ randomly selected time slots and plot the performance in dashed lines. It can be seen that the sum rate of the proposed approach is lower than that of the MSLNR-EP approach at the beginning because the ILCFs and BNCFs are actually random values due to the random initialization of the neural networks. Then, the sum rate of the proposed approach gradually increases as each BS optimizes its beamforming policy via the proposed DRL-based DDCBF framework. 
After about $4000$ time slots, the proposed approach finally converges with a close performance to the direct FP algorithm providing the upper bound and outperforms the other iterative optimization-based algorithms, namely, WMMSE-$50$RI, WMMSE-$10$RI, WMMSE, and closed-form FP. The closed-form FP and WMMSE algorithms realize similar performance as they are actually equivalent \cite{shen2018fractional}, and their performance can be improved by using more random initializations. Thus, we use the WMMSE algorithm to represent these two algorithms in the following comparisons for better readability. In this simulation setup, the ZG scheme can approach the WMMSE algorithm by only utilizing local CSI, which demonstrates its effectiveness. Moreover, these optimization-based algorithms jointly optimizing the beamformers of all BSs outperform the methods only utilizing local CSI, which demonstrates the performance gain of the centralized CBF algorithms. Similarly, the MSLNR-DDPG also converges with a sum rate a little higher than the MSLNR-EP method after about $4000$ time slots. 

\begin{figure}[tbp]
    \begin{center}
    \epsfxsize=0.5\linewidth
    \epsffile{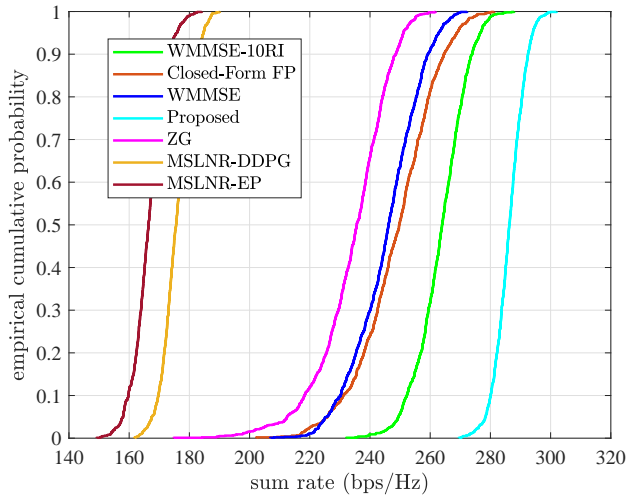}
    \caption{Downlink sum rate performance comparison of the proposed DRL-based DDCBF framework and the benchmarks.}\label{fig:perf_cmp_cdf}
    \end{center}
\end{figure}

After convergence, the cumulative distribution function (CDF) of the sum rates achieved by the proposed approach is shown in Fig. \ref{fig:perf_cmp_cdf}, and those of the benchmarks are also provided for comparison. Through the proposed DRL-based DDCBF framework, the BSs can learn the beamforming policies outperforming the iterative WMMSE algorithm. This is because the ILCFs at each BS are not necessarily the same in the proposed approach, while each BS has the same ILCFs in the WMMSE algorithm. In other words, the proposed approach makes it more possible for the BSs to find ILCFs and BNCFs providing better performance. Moreover, the MSLNR-DDPG scheme can also outperform the MSLNR-EP scheme, which shows the effectiveness of the proposed approach in only optimizing the power allocations. In addition, comparing the performance gain brought by optimizing the power allocations, optimizing the normalized beamformers can realize a larger performance improvement.

\blue{
In Table \ref{tab:overhead_comparison}, we also compare the proposed approach with the benchmarks in required information, communication overhead, and computational complexity (average computation time), and the simulation programs are executed with a personal computer with Intel i7-11700K inside. Thanks to the exploitation of neural networks, the average time required for each BS to obtain the beamformers in the proposed approach is only about $1.5$ ms. The average time required to obtain the beamformers via the WMMSE algorithm is approximately $40$ s because a large number of antennas lead to a large dimensional optimization problem of high computational complexity. Besides, the direct FP, which updates the beamformers by solving a convex problem in each iteration, requires more than $800$ s to find the solutions providing upper bound performance. It is worth noting that the computational complexity of the iterative ZG scheme is also very high since many iterations are required. Besides, we use the amount of information (quantified by the number of real scalar values) that needs to be collected from other cells to represent the communication overhead. It can be seen that the proposed approach can substantially reduce the communication overhead because it avoids an overhead increasing with $M$, e.g., the communication overhead of the proposed approach and the centralized optimization-based algorithms are respectively $928$ and $21504$.}

\begin{table*}[!t]
    \centering
    \caption{Comparison of the proposed approach and the considered benchmarks}
    \label{tab:overhead_comparison}
    \renewcommand{\arraystretch}{0.8}
    \resizebox{0.9\textwidth}{!}{\centering   
    \begin{tabular}{c||c|c|c}\hline
    Schemes & Required Information & Communication Overhead & \makecell{Computation Time\\ ($M=64, N=7, K=4$)} \\\hline\hline
    \makecell{MSLNR-EP} & \multirow{2}*{\makecell[c]{Local CSI}} & \multirow{2}*{0} & 0.7 ms\\\cline{1-1}\cline{4-4}
    \makecell{ZG} &  &  & 31.98 s\\\hline
    \makecell{WMMSE} & \multirow{3}*{\makecell[c]{Global CSI}} & \multirow{3}*{$2MN(N-1)K$} & 40.30 s\\\cline{1-1}\cline{4-4}
    \makecell{Closed-Form FP} &  &  & 38.84 s\\\cline{1-1}\cline{4-4}
    \makecell{Direct FP} &  &  & 815.11 s\\\hline
    \makecell{MSLNR-DDPG} & \multirow{2}*{\makecell[c]{Local CSI \& Historical \\information from other cells}} & \multirow{2}*{$(3N_c+1)UK^2+6UK$} & 1.3 ms\\\cline{1-1}\cline{4-4}
    \makecell[c]{\textbf{DDPG} (\textbf{proposed})} &  &  & 1.5 ms\\\hline
    \end{tabular}
    }
\end{table*}

\blue{
\subsection{Impacts of Handcraft Designs and Training Strategies}
}

\begin{figure}[htbp]
    \begin{center}
    \epsfxsize=0.5\linewidth
    \epsffile{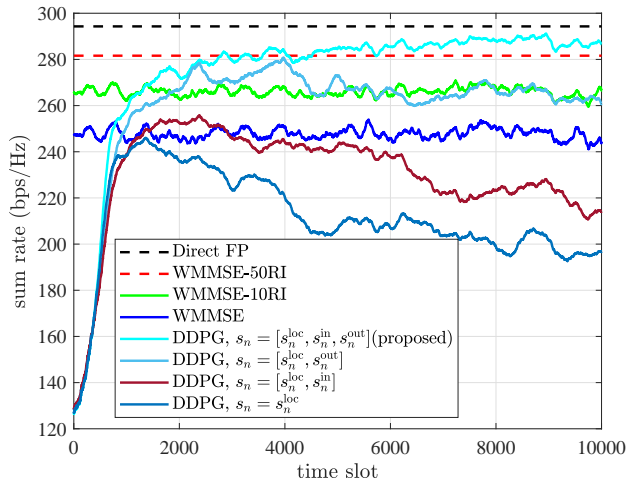}
    \caption{Downlink sum rate performance of the proposed DRL-based DDCBF framework for different state designs.}\label{fig:state_gain}
    \end{center}
\end{figure}
\blue{
Although the real-time global CSI requirement can be eliminated, each BS should still acquire some interference-related information as shown in the proposed handcraft state designs. To investigate the impacts of the handcraft designs, we compare the performance of the proposed DRL-based approach with different state designs, i.e., $s_{n}^{loc}$, $[s_{n}^{loc},s_{n}^{in}]$, and $[s_{n}^{loc},s_{n}^{out}]$. As shown in Fig. \ref{fig:state_gain}, the proposed approach with only the local information performs the worst and suffers a non-stationary convergence problem. By adding some interference-related information, namely, $s_{n}^{\rm in}$ or $s_{n}^{\rm out}$, the non-stationary convergence problem can be slightly alleviated. Due to the lack of sufficient information, it is difficult for the BSs to identify the wireless environment and learn an optimal beamforming policy. Besides, it can also be observed that the interfered UEs' information can realize a larger performance gain as compared to the interferers' information. This can be explained by the fact that the ICLFs depend more on the interference leakage to the interfered UEs, and thus the interfered UEs' information can help more. When both the interferer BSs' information and interfered information are involved in the agent's state, the proposed approach can realize a close performance to the direct FP algorithm.}

\begin{figure}[htbp]
    \begin{center}
    \epsfxsize=0.5\linewidth
    \epsffile{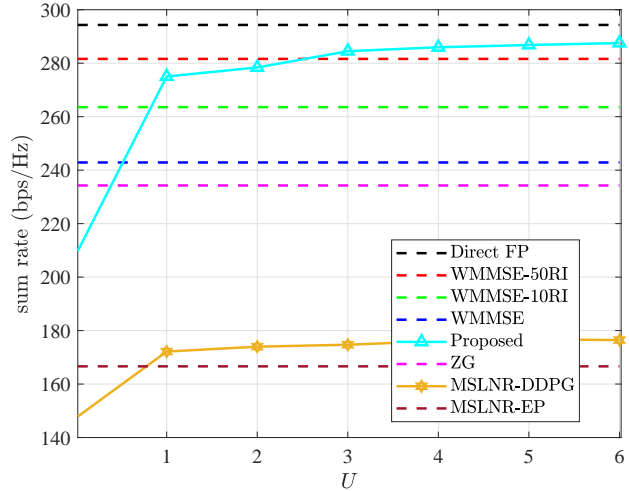}
    \caption{Downlink sum rate performance of the proposed DRL-based DDCBF framework for different amounts of historical inter-cell information.}\label{fig:perf_U}
    \end{center}
\end{figure}

On the other hand, as the historical information from other cells can help the BS identify the wireless environment and learn better policy, one may want to know the impact of the amount of historical inter-cell information on the performance of the proposed framework. We recall that the amount of historical inter-cell information mainly depends on the parameter $U$. In Fig. \ref{fig:perf_U}, we investigate the performance of the proposed framework for different $U$'s. It can be seen that the performance of the proposed approach improves as $U$ increases since more historical inter-cell information allows the BS to learn more from the wireless environment and make decisions more accurately. Besides, the performance improvement brought by the increase of $U$ becomes small when $U\geq 4$, which means that $U=4$ is sufficient for the proposed approach to realize a near-optimal performance. Besides, similar phenomenons can be observed by comparing the performance of the MSLNR-DDPG scheme with that of the MSLNR-EP scheme.

\begin{figure}[tbp]
    \begin{center}
    \epsfxsize=0.5\linewidth
    \epsffile{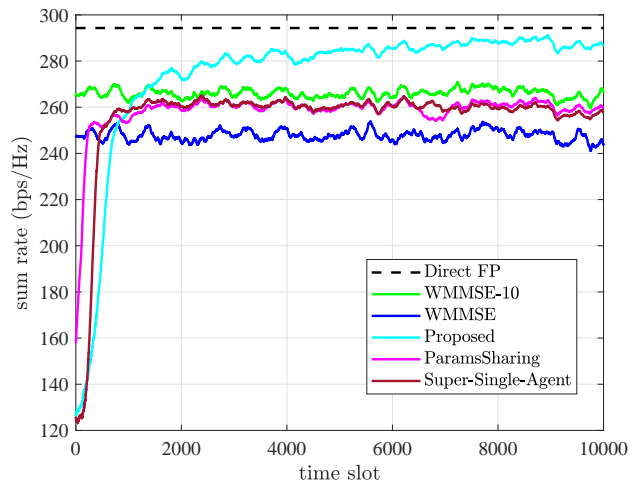}
    \caption{Comparison of downlink sum rate performance for different training strategies.}\label{fig:training_manner}
    \end{center}
\end{figure}

\blue{
In addition, we also investigate the impacts of different training strategies on the considered multi-agent system. Two alternatives in the existing literature are considered, i.e., parameter sharing and super-single-agent methods \cite{gupta2017cooperative}, where the global replay buffer is invoked. Specifically, the parameter-sharing method allows each BS to share a global policy and the policy is trained with the experiences of all agents in a global replay buffer. The super-single-agent method jointly trains all the agents in a centralized manner by assuming a joint model for the actions and observations of all the agents and thus can be regarded as training a super-single-agent with state space $s=[s_1, \cdots, s_{N}]$, action space $a=[a_1, \cdots, a_{N}]$, and a reward function set as the sum rate of all users. In Fig. \ref{fig:training_manner}, we provide the performance of these two alternative training strategies. Both of the two schemes can outperform the WMMSE algorithm and show faster convergences, but their performance is slightly lower than the proposed approach. This can be explained as follows. With parameter sharing, the global policy can be trained more efficiently, however, the distinctions between the optimal policies of the BSs are neglected. The super-single-agent scheme leads to a much larger action space, which makes it harder to find the optimal policy via limited trial-and-error interactions. In the proposed approach, we utilize the distributed reward function to realize a decentralized training approach, and the distributed reward function can also be regarded as a net gain of the impact of the BS's action on the objective function. With this design, the BS's action can be evaluated efficiently, and the proposed approach can finally realize a performance close to the direct FP algorithm.
}

\subsection{Evaluation of Proposed DRL-based DDCBF Framework}

\begin{figure}[tbp]
    \begin{center}
    \epsfxsize=0.5\linewidth
    \epsffile{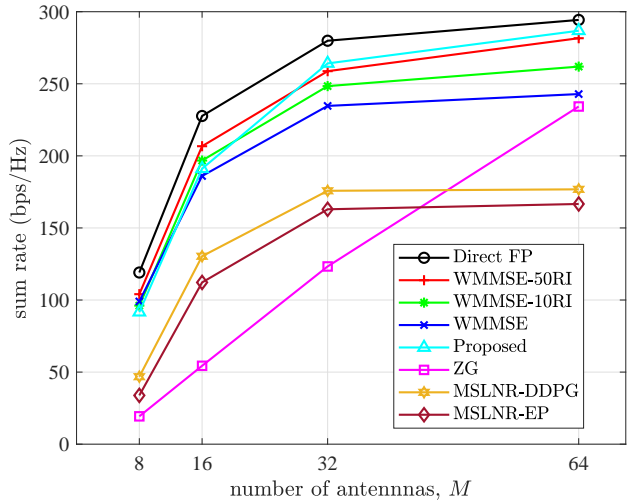}
    \caption{Downlink sum rate performance comparison for different numbers of antennas at each BS, where the number of users per cell is set as $K=4$.}\label{fig:perf_ants_3}
    \end{center}
\end{figure}

Further, we investigate the performance of the proposed approach in different simulation setups, i.e., different numbers of antennas at each BS and different numbers of users per cell. In Fig. \ref{fig:perf_ants_3}, we show the sum rates achieved by all the schemes for different numbers of antennas at each BS, namely, $M$. As $M$ becomes larger, all the schemes can achieve higher sum rates because the BS can realize higher array gains and more efficient interference management by exploiting more antennas. Besides, the performance improvement brought by the increase of $M$ become smaller as $M$ becomes larger in most schemes. This is because the achievable rate is a concave function of SINR, and therefore the improvement in the achievable rate can be regarded as a marginal utility of SINR. With larger $M$, higher array gains and more efficient interference management can be realized to improve the SINR performance, and the improvement in the achievable rate decreases according to the marginal utility law \cite{frank2010microeconomics}. \blue{When $M=8$, we can see that the proposed approach realizes a performance slightly lower than the WMMSE algorithm. When $M=64$, the proposed approach can outperform WMMSE-$50$RI, which shows the effectiveness of the proposed approach in a large $M$ regime. This is because the exploitation of expert knowledge can substantially reduce the number of parameters to determine, and the proposed approach can find better solutions for such large-dimensional optimization problems. It is also more difficult for the optimization-based algorithm to find near-optimal solutions to the large dimensional optimization problems since there may exist more local optimums.} The effectiveness of the proposed approach in obtaining the near-optimal beamformers is further validated. Although the ZG-based method can realize a comparable performance to the WMMSE algorithm, its performance is upper bounded by that of the WMMSE algorithm \cite{choi2012distributed}. \blue{Moreover, the effectiveness of the ZG-based method vanishes in the small $M$ regime, which means that there exists a large performance loss in decoupling the original problem into the sub-problems involving only local CSI when the local CSI are spatially correlated}.

\begin{figure}[tbp]
    \begin{center}
    \epsfxsize=0.5\linewidth
    \epsffile{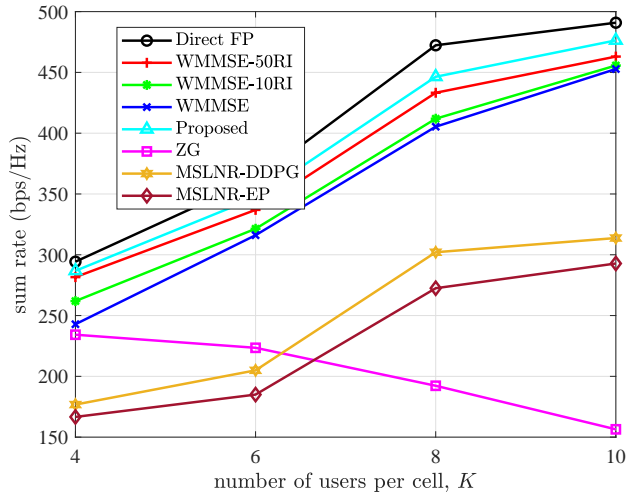}
    \caption{Downlink sum rate performance comparison for different numbers of users per cell, where the number of antennas at each BS is set as $M=64$.}\label{fig:perf_ue_3}
    \end{center}
\end{figure}

In some sense, the marginal effect of the sum rate shown in Fig. \ref{fig:perf_ants_3} is due to the limited number of users in the network. Hence, we also investigate the performance of the sum rate of all the schemes for different numbers of users per cell. As shown in Fig. \ref{fig:perf_ue_3}, except for the ZG-based algorithm, the sum rates achieved by the other schemes become larger with the increase in the number of users per cell. \blue{The degrading performance of the ZG-based scheme with the increase of $K$ is again because the local CSI becomes spatially correlated for larger $K$.} Besides, the proposed approach can always outperform the WMMSE algorithm for different $K$'s since $M=64$ is fixed. Thus, the proposed DRL-based DDCBF framework can provide us with a more efficient method to obtain the near-optimal beamformers in the considered massive MIMO mobile cellular network in comparison with the WMMSE algorithm. Moreover, the performance improvement achieved by the CBF technique, i.e., the gap between the sum rate of the WMMSE algorithm (or the proposed approach) and that of the MSNR-EP method becomes larger as $K$ increases. This can be explained by the fact that the interference in the network becomes more severe as the number of users increases, and therefore the CBF technique can achieve a more notable performance improvement by managing the interference efficiently.

\section{Conclusions}\label{sec:conclusion}

In this work, we have studied a dynamic coordinated beamforming problem in massive MIMO cellular networks. As the conventional optimization-based algorithms are not practically viable due to the high computational complexity and the real-time global CSI requirement, we have proposed a DRL-based distributed dynamic coordinated beamforming framework leveraging expert knowledge, i.e., a known solution structure observed from the iterative procedure of the WMMSE algorithm. In particular, the proposed approach allows each BS to obtain near-optimal beamformers with only local CSI and some historical information from other cells. Besides, the computational complexity for each BS to obtain the beamformers is considerably low because of the exploitation of neural networks, which makes it possible for the BS to calculate the beamformers quickly once the channel varies. Finally, we have provided extensive numerical simulations to evaluate the proposed DRL-based approach, and the results have shown that the proposed approach can realize a close performance to the existing upper bound optimization-based algorithm.

\bibliographystyle{IEEEtran}
\bibliography{IEEEabrv, ref_MADRL4DDCBF}

\end{document}